\documentclass[twocolumn,english,prl,showpacs,superscriptaddress]{revtex4-1}
\usepackage[T1]{fontenc}
\usepackage[latin9]{inputenc}
\usepackage{amstext}
\usepackage{graphicx}
\usepackage{babel}

\begin{document}

\title{Giant Forward Scattering Asymmetry and Anomalous Tunnel Hall effect at Spin-Orbit-and Exchange-Split Interfaces.}
\author{T. Huong Dang}
\affiliation{Laboratoire des Solides Irradi\'es, Ecole Polytechnique, CNRS UMR 7642, and CEA-DSM-IRAMIS, University Paris Saclay, 91128 Palaiseau cedex, France}
\author{H. Jaffr\`es}
\affiliation{Unit\'e Mixte de Physique CNRS-Thales, 1, Av. Augustin Fresnel 91767, Palaiseau \\ and Universit\'e Paris-Sud 91405, Orsay, France}
\author{T.~L. Hoai Nguyen}
\affiliation{Institute of Physics, VAST, 10 Daotan, Badinh, Hanoi, Vietnam.}
\author{H.-J. Drouhin}
\affiliation{Laboratoire des Solides Irradi\'es, Ecole Polytechnique, CNRS UMR 7642, and CEA-DSM-IRAMIS, University Paris Saclay, 91128 Palaiseau cedex, France}

\begin{abstract}
We report on theoretical investigations of scattering asymmetry \textit{vs.} incidence of carriers through exchange barriers and magnetic tunnel junctions made of semiconductors involving spin-orbit interaction. By an analytical $2\times 2$ spin model, we show that, when Dresselhaus interaction is included in the conduction band of antiparallel magnetized electrodes, the electrons can undergo a large difference of transmission depending on the sign of their incident in-plane wavevector. In particular, the transmission is fully quenched at some points of the Brillouin zone for specific in-plane wavevectors and not for the opposite. Moreover, it is universally scaled by a unique function independent of the spin-orbit strength. This particular feature is reproduced by a $14\times 14$ band $\mathbf{k}\cdot \mathbf{p}$ model showing, in addition, corresponding effects in the valence band and highlighting the robustness of the effect, which even persists for a single magnetic electrode. Upon tunneling, electrons undergo an asymmetrical deflection which results in the occurrence of a transverse current, giving rise to a so-called Tunnel Hall Effect.
\end{abstract}


\date{\today}
\maketitle

The interplay between particle spin and orbital motion  is currently at the basis of new functionalities. These generally require efficient spin-current injection at magnetic-nonmagnetic interfaces, efficient spin-transfer torque, and possibly efficient spin Hall effect with heavy materials~\cite{liu2012,miron2010,garello2013,rojas2014} for magnetic commutation. Spin-orbit interaction ($SOI$) at an interface with broken inversion symmetry may lead to the observation of Rashba-split states~\cite{bychkov1984,bihlmayer2006,krupin2009,gambardella2011} which can be used to convert a perpendicular spin-current into a lateral charge current by inverse Edelstein effect ($IEE$)~\cite{rojas2013,viret2015}. In that context, investigations of $SOI$ in solids and at interfaces is of prime importance for basic physics and today's technology.

In this paper, we study unconventional quantum effects resulting in giant transport asymmetry for electrons or holes in structures composed of exchange and spin-orbit-split electrodes made of III-V semiconductors with antiparallel ($AP$) magnetizations, possibly separated by thin tunnel barriers. The symmetry of the structure allows a transmission difference~\textit{vs.} carrier incidence near interfaces with respect to the reflection plane defined by the magnetization and the surface normal. This quantum process departs from the effect of a beam deviation by the Lorentz force due to the action of a local magnetic field in the barrier~\cite{alekseev2010} and from spin-filtering effects occurring in non centro-symmetric structures~\cite{perel2003,perel2004}. Unlike the latter, the effect we propose requires the simultaneous action of both in-plane and out-of plane spin-orbit fields for promoting transport spin asymmetry. In order to address the issue in a simple way, we first consider a heterojunction made of two identical magnetic semiconductors of zinc-blende symmetry, with opposite in-plane magnetizations: this structure (Fig.~1) constitutes an ideal exchange step and a paradigm for exchange-engineered heterostructures, similarly to symmetrical spin valves in giant magnetoresistance~\cite{vanKempen1987,valetfert1993}. Indeed, due to the axial character of the magnetization, the $AP$ configuration breaks the symmetry with respect to the reflection plane (in Fig.~1 the reflection plane is the $xz$ plane), and also some possible rotation and time conjugation invariances existing in the parallel $(PA)$ magnetic arrangement~\cite{note1}. The result is that two waves with opposite in-plane wavevectors $\mathbf{k}_\parallel$ may be differently transmitted in amplitude.

\begin{figure}
\includegraphics[height=6cm]{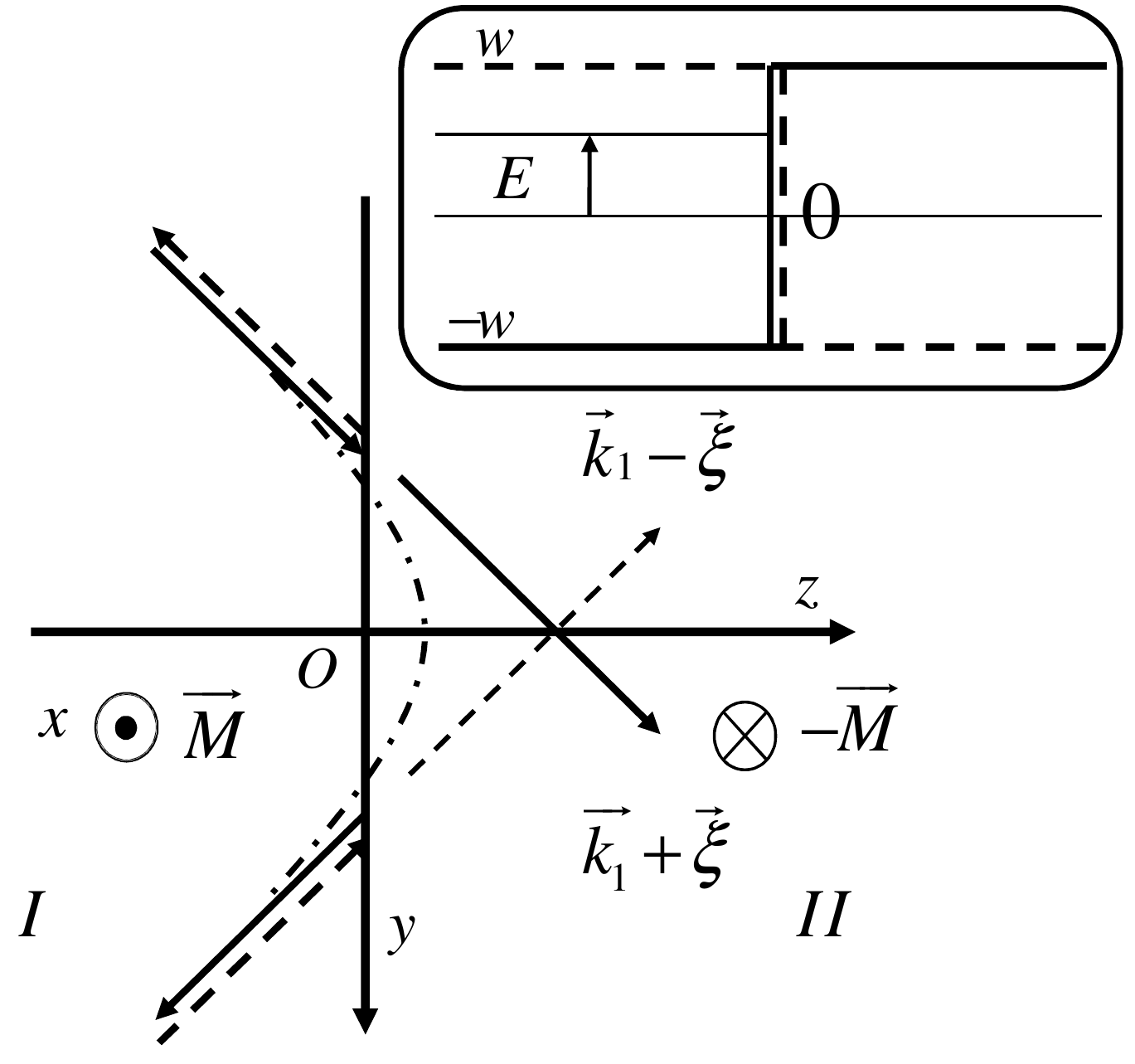}
\caption{Scheme of transmission process at an exchange-$SOI$ step with $AP$ magnetizations $\mathbf{M}$ and $-\mathbf{M}$ along $x$. The propagation direction of carriers (straight arrow) is along $z$ with propagative wavevector $k_{1}$ whereas the in-plane incident component $+ \xi $ (heavy line) or $- \xi$ (dashed line) is along $y$; $\mathbf{xyz}$ forms a direct frame. The dash-dot curve denotes the evanescent waves, either reflected or transmitted. Carriers with $+\xi $ in-plane wavevector component are more easily transmitted than those carrying $-\xi .$(Top right inset): Energy profile of the exchange step;  $E$ is the longitudinal kinetic energy along $z$ and $2w$ is the exchange splitting in the magnetic materials.}\label{fig1}
\end{figure}

We first consider Dresselhaus interaction in the conduction band of bulk materials~\cite{Dresselhaus}. Hereafter, we refer the structure to the $x, y, z$ cubic axes (unit vectors $\widehat{x},\widehat{y},\widehat{z}$) and assume that electron transport occurs along the $z$ axis, whereas the magnetization lies along $x$. We study the transmission asymmetry when the wavevector component along $y$ is changed from $\xi $ to $-\xi $. Electrons are injected from the first conduction band of material $I$ to the left ($\epsilon =1$) into the first conduction band of material $II$ to the right ($\epsilon =-1$). Then, the relevant $2\times 2$ Hamiltonians respectively write:
\begin{eqnarray}
\widehat{H}_{I,II} &=&\gamma _{c}\left( k^{2}+\xi ^{2}\right) \widehat{I}+w\mathbf{m}\cdot \widehat{\mathbf{\sigma }}+\left( \widehat{\gamma }\mathbf{\chi }\right) \cdot \widehat{\mathbf{\sigma }}  \nonumber \\
&=&\left(
\begin{array}{cc}
\gamma _{c}(k^{2}+\xi ^{2})-\widetilde{\gamma }\xi ^{2}k & -i\gamma \xi k^{2}+\epsilon w \\
i\gamma \xi k^{2}+\epsilon w & \gamma _{c}(k^{2}+\xi ^{2})+\widetilde{\gamma
}\xi ^{2}k%
\end{array}%
\right)   \label{H}
\end{eqnarray}%
where $(0,\xi, k)$ is the electron wavevector; $\widehat{I}$ is the identity matrix,\ $\gamma _{c}$ accounts for the conduction effective mass, $\mathbf{m}$ is the unit magnetization vector, $2w$ the exchange splitting (assumed to be positive), $\widehat{\mathbf{\sigma }}$ the Pauli operator, and $\mathbf{\chi}=\left[ 0,\xi k^{2},-\xi^{2}k\right]$ the D'yakonov-Perel' (DP) internal field responsible for the spin splitting~\cite{DyaPer,Dresselhaus}. For the subsequent discussion, we introduce the tensor $\widehat{\gamma }=\left( \gamma_{i}\delta _{ij}\right) $ which characterizes the DP-field strength, with $\gamma _{x}=\gamma _{y}=\gamma $, $\gamma _{z}=\widetilde{\gamma }$, and $\delta _{ij}$ the Kronecker symbol. We consider the two cases $ \widetilde{\gamma }=\gamma $ and $\widetilde{\gamma }=0$, switching on and off the $\xi ^{2}$ diagonal perturbation.

The two energies in the exchange and spin-orbit-split subbands are given by $\mathcal{E}_{1}=$ $\gamma _{c}\left(k_{1}^{2}+\xi ^{2}\right) -w$ and $\mathcal{E}_{2}=\gamma _{c}(k_{2}^{2}+\xi^{2})+w$, where $k_{1}$ $\left( k_{2}\right) $ is the $z$-component of the wavevector in the lower (upper) subband. These expressions are correct up to
first order in $\gamma $ provided  $\left\vert \widetilde{\gamma }\xi ^{2}k/w\right\vert <<1$\ \ and\ \ $\left\vert \gamma \xi k^{2}/w\right\vert <<1$, where$\ k=k_{1}$ or $k=k_{2}$. The respective eigenvectors  write:
\begin{eqnarray}
\mathbf{u}_{\epsilon ,1}\left( \xi ,k_{1}\right)  &=&\left[ 1-2\epsilon i\mu k_{1}^{2},-\epsilon \left( 1-2\widetilde{\mu }\xi k_{1}\right) \right] /\sqrt{2}\text{,}  \label{u1} \\
\mathbf{u}_{\epsilon ,2}\left( \xi ,k_{2}\right)  &=&\left[ 1-2\epsilon i\mu k_{2}^{2},\epsilon \left( 1+2\widetilde{\mu }\xi k_{2}\right) \right] /\sqrt{2}\text{,}  \label{u}
\end{eqnarray}%
where $\mu =\gamma \xi /(2w)$ and $\widetilde{\mu }=\widetilde{\gamma }\xi /(2w)$. Note that the norm of $\mathbf{u}_{\epsilon ,\ell }$ ($\ell $ $=1$ or $2$) only involves even powers of $\xi $ likewise the direct overlap $\left\vert \left\langle \mathbf{u}_{\epsilon ,\ell }|\mathbf{u}_{-\epsilon ,\ell }\right\rangle\right\vert ^{2}$ between incoming and outgoing states, so that no $\pm \xi$ transmission asymmetry can be expected in usual tunneling models, \textit{e.g.} based on interface density of states~\cite{bardeen1961,julliere1975,slonczewski1989}. The asymmetry appears in full-quantum treatments involving matching conditions at interfaces and may be correctly described by embedding methods~\cite{wortmann2002}.

The corresponding wavefunctions in Regions $I$ and $II$ can be written in a compact form:
\begin{eqnarray*}
\Psi _{I}\left( z\right)  &=&\alpha \mathbf{u}_{1,2}\left( \xi ,k_{2}\right)\;e^{ik_{2}z}+\beta \mathbf{u}_{1,1}\left( \xi ,k_{1}\right) \;e^{ik_{1}z} \\
&&+A\mathbf{u}_{1,2}\left( \xi ,-k_{2}\right) \;e^{-ik_{2}z}+B\mathbf{u}_{1,1}\left( \xi ,-k_{1}\right) \;e^{-ik_{1}z}\text{,}
\end{eqnarray*}%
\begin{equation}
\Psi _{II}\left( z\right) =C\mathbf{u}_{-1,1}\left( \xi ,k_{1}\right)\;e^{ik_{1}z}+D\mathbf{u}_{-1,2}\left( \xi ,k_{2}\right) \;e^{ik_{2}z}\text{,}  \label{wavefunction}
\end{equation}%
where the $\alpha $ and $\beta $ (resp. $A$ and $B$) amplitudes stand for incident waves (resp. reflected waves) in Region $I$, and $C$ and $D$ for transmitted waves in Region $II$. Because $k_{\parallel }$ is conserved in the transport process, we are dealing with states with the same \textit{longitudinal} kinetic energy $E$ along $z$ and a total kinetic energy $\mathcal{E}=E+\gamma _{c}\xi ^{2}$. The proper matching conditions are the continuity of the wavefunction and of the current wave $\widehat{J}\Psi_{I,II}=\left(1/\hbar \right) (\partial \widehat{H}_{I,II}/\partial k) \Psi_{I,II}$ because $\widehat{H}_{I,II}$ contains no more than quadratic $k$ terms~\cite{petukhov2002,elsen2007bis,Hoai2009,Drouhin2011,BottPRB2012} and because $\hat{\gamma}/\gamma _{c}$ is continuous.
\begin{figure}[tbp]
\includegraphics[height=12cm]{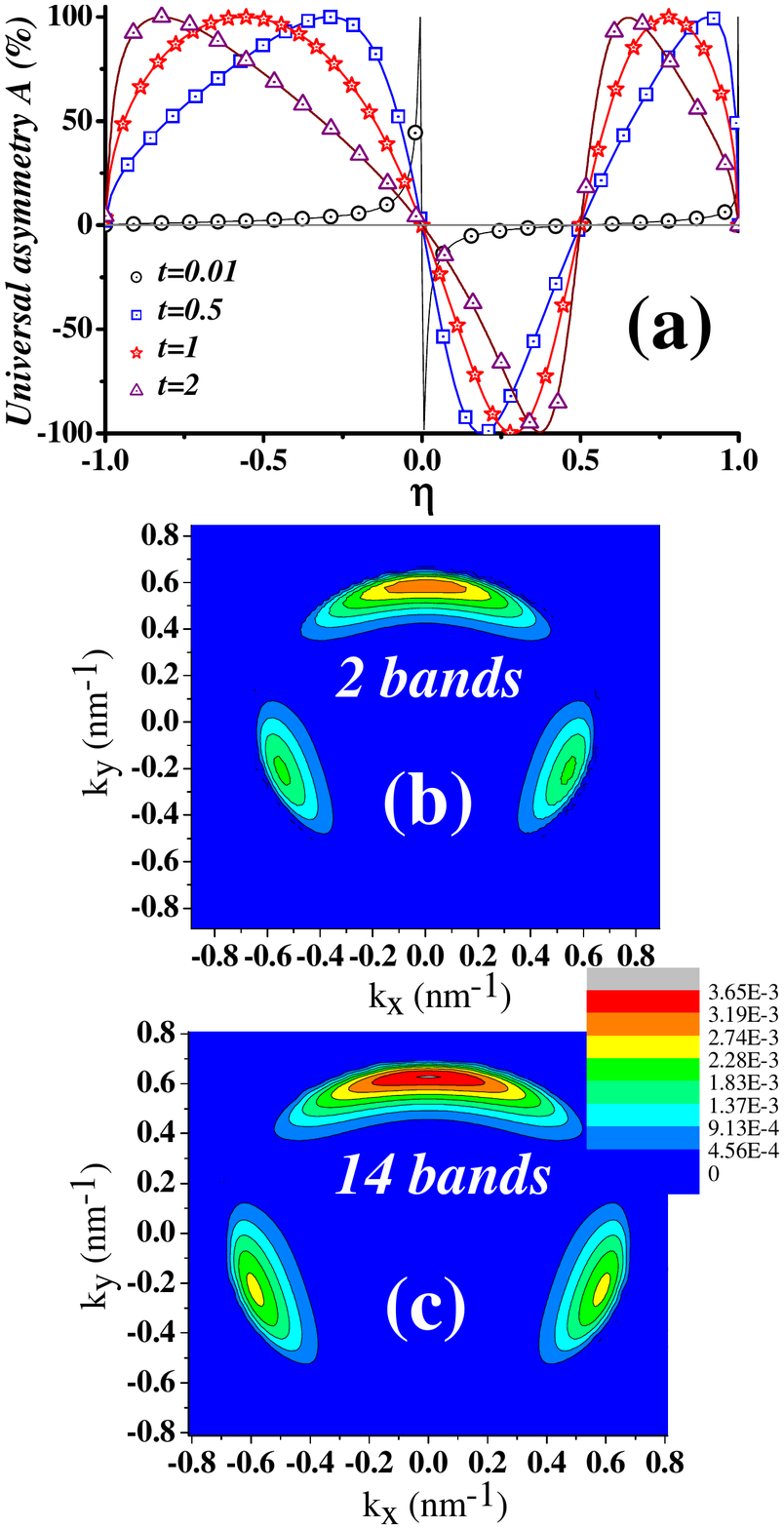}
\caption{(a) Universal asymmetry coefficient $\mathcal{A}$ \textit{vs.} reduced energy $\protect\eta =E/w$ obtained for different values of $t=\protect\xi /K$\ [$t=0.01$ (black; circles), $t=0.5$ (blue; squares), $t=1$ (red; stars), and $t=2$ (purple; triangles) by $2$-band analytical (full line) and numerical (symbols) calculations. Two-dimensional map of the transmission coefficient $T$ in $2\times 2$ (b) and $14\times 14$ (c) $\mathbf{k}\cdot \mathbf{p}$ band models for the exchange-$SOI$ step schematized in Fig.~1; the parameters are: exchange energy $2w=0.3$ eV, \textit{total} kinetic energy $\mathcal{E}=0.08$ eV counted from the middle of the conduction step, and $\protect\gamma =-24$~eV~\AA $^{\text{3}}$; band parameters of the $14$-band $\mathbf{k}\cdot \mathbf{p}$ model taken from Ref.~\cite{jancu2005}}
\end{figure}
The average transmission coefficient $T\left( \xi ,k_{1},k_{2}\right) $ upon positive and negative incidences is related to the amplitude of the transmitted wave $C\left( \xi,k_{1},k_{2}\right) $ calculated with the initial conditions $\alpha =0$ and $\beta =1$ through:
\begin{equation}
T\left( \xi ,k_{1},k_{2}\right) =\frac{\left\vert C\left( \xi,k_{1},k_{2}\right) \right\vert ^{2}+\left\vert C\left( -\xi,k_{1},k_{2}\right) \right\vert ^{2}}{2}\text{,}  \label{Tr1}
\end{equation}%
and we define the transmission asymmetry as:
\begin{equation}
\text{\ \ \ \ }\mathcal{A}\left( \xi ,k_{1},k_{2}\right) \ =\frac{\left\vert C\left( \xi ,k_{1},k_{2}\right) \right\vert ^{2}-\left\vert C\left( -\xi,k_{1},k_{2}\right) \right\vert ^{2}}{\left\vert C\left( \xi,k_{1},k_{2}\right) \right\vert ^{2}+\left\vert C\left( -\xi,k_{1},k_{2}\right) \right\vert ^{2}}.  \label{As}
\end{equation}

It can be checked that, when $\widetilde{\gamma}=0$, $\mathcal{A}\left( \xi,k_{1},k_{2}\right) $ vanishes if $\alpha $ and $\beta $ are real, which is a non trivial result. The transmission of a pure up-spin incident electron into a pure down-spin state is only possible under oblique incidence \textit{via} $SOI$ which introduces off-diagonal matrix elements. Moreover, a non-vanishing diagonal part of $SOI$ is necessary to obtain a non-zero asymmetry although the $z$ component of the DP field along $z$ does not depend on the sign of $k_{\parallel}$~\cite{note4}. Then, from now on, we take $\widetilde{\gamma }=\gamma $. The wavevector $k_{1}$ in the lower subband has to be real so that we can define $K=k_{1}>0$. We introduce the parameter $\lambda>0$ with $k_{2}=i\lambda K$, the reduced longitudinal energy $\eta=E/w=\left( 1-\lambda^{2}\right) /\left( 1+\lambda ^{2}\right)$, and the incidence parameter $t=\xi /K$. One obtains:

\begin{equation}
C\left( \xi ,K,\lambda \right) =\left(\frac{\gamma K^{2}\xi}{w} \right)  \frac{\left[ \left( \xi /K\right) \left( 3\lambda ^{2}-1\right) +2\lambda \left( \lambda ^{2}-1\right) \right] }{\left( \lambda -i\right) ^{2}}\text{.}
\label{C}
\end{equation}

From Eq.~\ref{C}, it can be checked that $\mathcal{A}\left(\xi,k_{1},k_{2}\right) =0$ if $\lambda $ is purely imaginary; the asymmetry appears when the lower-energy band carries a propagative state whereas the upper one acts as a barrier sustaining an evanescent state. Transport is then described in a two-$k$-channel model, a propagative channel ($k_{1}$) and an evanescent channel ($k_{2}$). One obtains:
\begin{equation}
T\left( t,\eta \right)\ =\frac{\gamma ^{2}}{\gamma _{c}^{3}}wt^{2}\left(1+\eta \right) ^{2}\left[4\eta ^{2}\left(1-\eta \right)+t^2 (1+\eta)(2\eta -1)^{2} \right] \text{,}  \label{T}
\end{equation}

and
\begin{equation}
\mathcal{A}\left( t,\eta \right) \ =\frac{4t \eta \sqrt{1-\eta ^{2}}\left(2\eta -1\right) }{4\eta ^{2}\left( 1-\eta \right) +t^{2}\left( 1+\eta \right) \left( 2\eta -1\right) ^{2}}\text{.}  \label{A}
\end{equation}

Equation~\ref{T} emphasizes the increase of $T\left( t,\eta \right)$  with $t$ and $\gamma $. The range of validity defined above can be written $\left\vert t^{2}\left(\gamma K^{3}/\gamma_{c}K^{2}\right) \right\vert <<1$, a condition easily fulfilled.

The analytical asymmetry $\mathcal{A}$ is plotted in Fig.~2a for several values of $t$ (full lines), where the symbols refer to the $2\times 2$ numerical calculations, showing an excellent agreement. It can be seen that the curves related to $t$ and $t^{'}=1/t$ are located at almost symmetrical positions with respect to the $t=1$ curve. They admit four zeros in the energy range considered: \textit{i}) two at the two ends of the energy step when either the propagative or the evanescent state disappears and \textit{ii}) one in the middle of the energy barrier and one for $\eta=3/4$ which is particular to Dresselhaus interaction. It is a remarkable result that $\mathcal{A}\left( t,\eta \right) $ does not depend either on the material parameters or on the sign of $\gamma$, thus conferring to $\mathcal{A}$ a universal character. Reversing the magnetization (changing $w$ into $-w$) makes transport occur in the $k_{2}$ channel leading to a change of $\mathcal{A}\left(t,\eta \right)\ $ into $-\mathcal{A}\left( t,\eta \right) $~\cite{note2}. An important result is that $\mathcal{A}$ is positive when $\left( \mathbf{m},\mathbf{\xi },\mathbf{k}\right) $ forms a direct frame and negative otherwise. Another striking feature is that an arbitrarily small perturbation is able to produce a 100\% transport asymmetry \textit{i.e.}, a total quenching of transmission. Fig.~2b-2c display the 2-dimensional map of the electron transmission at a given total energy in the reciprocal space calculated using both a $2\times 2$ effective Hamiltonian (Fig.~2b) and a full $14\times 14$ band $\mathbf{k}\cdot \mathbf{p}$ treatment (Fig.~2c) involving odd-potential coupling terms $P^{^{\prime }}$ and $\Delta ^{^{\prime }}$~\cite{jancu2005,cardona1988,pfeffer1991}. These calculations are based on the multiband transfer matrix technique~\cite{petukhov2002, elsen2007bis,note3}. We have checked that transport asymmetry also arises for a tunnel junction where a thin tunneling barrier is inserted between the two magnetic layers. Tailoring more complicated structures which involve resonant tunneling effects allows one to obtain much higher transmission (a fraction of unity) while keeping the same magnitude for $\mathcal{A}$~\cite{noteSM}.

We can calculate the transmitted current, $\mathbf{J}\left[t,\eta\right]=\mathbf{J}_{\xi}\left[\Psi_{II}\left(z\right)  \right]  +\mathbf{J}_{-\xi}\left[\Psi_{II}\left(
z\right)\right]$, originating from incident waves of equal amplitude with opposite $k_{\parallel}$. To the lowest order in $\gamma$, we find
\begin{equation}
\mathbf{J}_{y,z}\left[  t,\eta\right]  =\frac{4\left(  \gamma_{c}w\right)
^{1/2}}{\hbar}\left(  1+\eta\right)  ^{1/2}T\left(  t,\eta\right)  \left[
\mathcal{A}\left(  t,\eta\right)  t\widehat{y}\mathbf{+}\widehat{z}\right]
\label{current}
\end{equation}
Thus, the asymmetrical transmission gives rise to a transverse momentum and then to a tunneling surface current (per unit length) $j_y=J_y \times \ell_{mfp}$ ($\ell_{mfp}$ is the electron mean free path) which can lead to an anomalous Tunnel Hall Effect ($THE$) under steady state regime. This effect could be experimentally investigated at a scale where the thickness of the channel collecting the current is comparable to $\ell_{mfp}$, \textit{i.e.}, not exceeding a few nm~\cite{note5}. The ratio of the (surface) transverse to the longitudinal current $j_{_{y}}\left[t,\eta\right]/J_{z}\left[t,\eta\right]=t\mathcal{A}\left(t,\eta\right)\ell_{mfp}$ then defines the $THE$ length in the spirit of a recent work dealing with $IEE$ phenomenon~\cite{rojas2013,viret2015}. To gain some numerical insight, an incident beam in Region I with a given angular dispersion with respect to the $z$ axis gives rise, after angular averaging at a fixed given energy~\cite{note6}, to a $THE$ length $t \mathcal{A} \ell_{mfp}$ of the order of $\ell_{mfp}$ for a beam deviation as large as 45°~\cite{noteSM}.

What happens for the valence bands simply described within the 6-band Luttinger $\mathbf{k}\cdot \mathbf{p}$ effective Hamiltonian~\cite{luttinger1956}? The results are shown in Fig.~3; we have checked that the $14$-band model provides similar data with $P^{^{\prime}}=0$ and $\Delta^{^{\prime}}=0$, surprisingly showing that the absence of inversion symmetry is not a key feature in the valence band. The lower curve displays the asymmetry $\mathcal{A}$ \textit{vs.} hole energy $\mathcal{E}$ in the case of a 3-nm-thick tunnel barrier. The energy range covers the valence spin subbands, namely, starting from the highest energy, the up-spin heavy (light)-hole band $HH\uparrow$ ($LH\uparrow $), the down-spin light (heavy)-hole band $LH\downarrow $ ($HH\downarrow $), and the up (down)-spin split-off band $SO\uparrow $($SO\downarrow $). We refer to points (1) to (6) marked by vertical arrows in the following discussion. Here, the energy of the $HH\uparrow $ [$HH\downarrow $] maximum, corresponds to $0.15$~eV [$-0.15$~eV], the energy origin being taken at the top of the valence band of the non-magnetic material, and is indicated by point (1) [(4)]. Correspondingly, one observes an almost fully negative transmission asymmetry in this energy range for predominant majority up-spin injection as far as $HH\downarrow $ does not contribute to the current. At more negative energy [$\mathcal{E}<-0.15$~eV: point (4)], a sign change of $\mathcal{A}$ occurs at the onset of $HH\downarrow $ (in the upper left inset, see the step in the transmission coefficient, which reaches almost +50\%); $\mathcal{A}$ remains positive after crossing $SO\uparrow $ [point (5)] before turning negative again once crossing $SO\downarrow $ [point (6)]. Note that $\mathcal{A}$ changes sign two times at characteristic energy points corresponding to a sign change of the injected particle spin. We have performed similar calculation for a simple contact ($d=0$; right upper inset in Fig.~3, black curve). Remarkably, $\mathcal{A}$, although smaller, keeps the same trends as for the 3-nm tunnel junction, except for a change of sign, showing a subtle dependence of the exchange coupling on the barrier thickness. Without tunnel junctions, $\mathcal{A}$ abruptly disappears as soon as $SO\downarrow $ contributes to tunneling [circle region] \textit{i.e.}, when evanescent states disappear. In the case of tunnel junction, $\mathcal{A}$, although small, subsists in this energy range and this should be related to the evanescent character of the wavefunction in the barrier.

\begin{figure}[tbp]
\includegraphics[height=6.5cm]{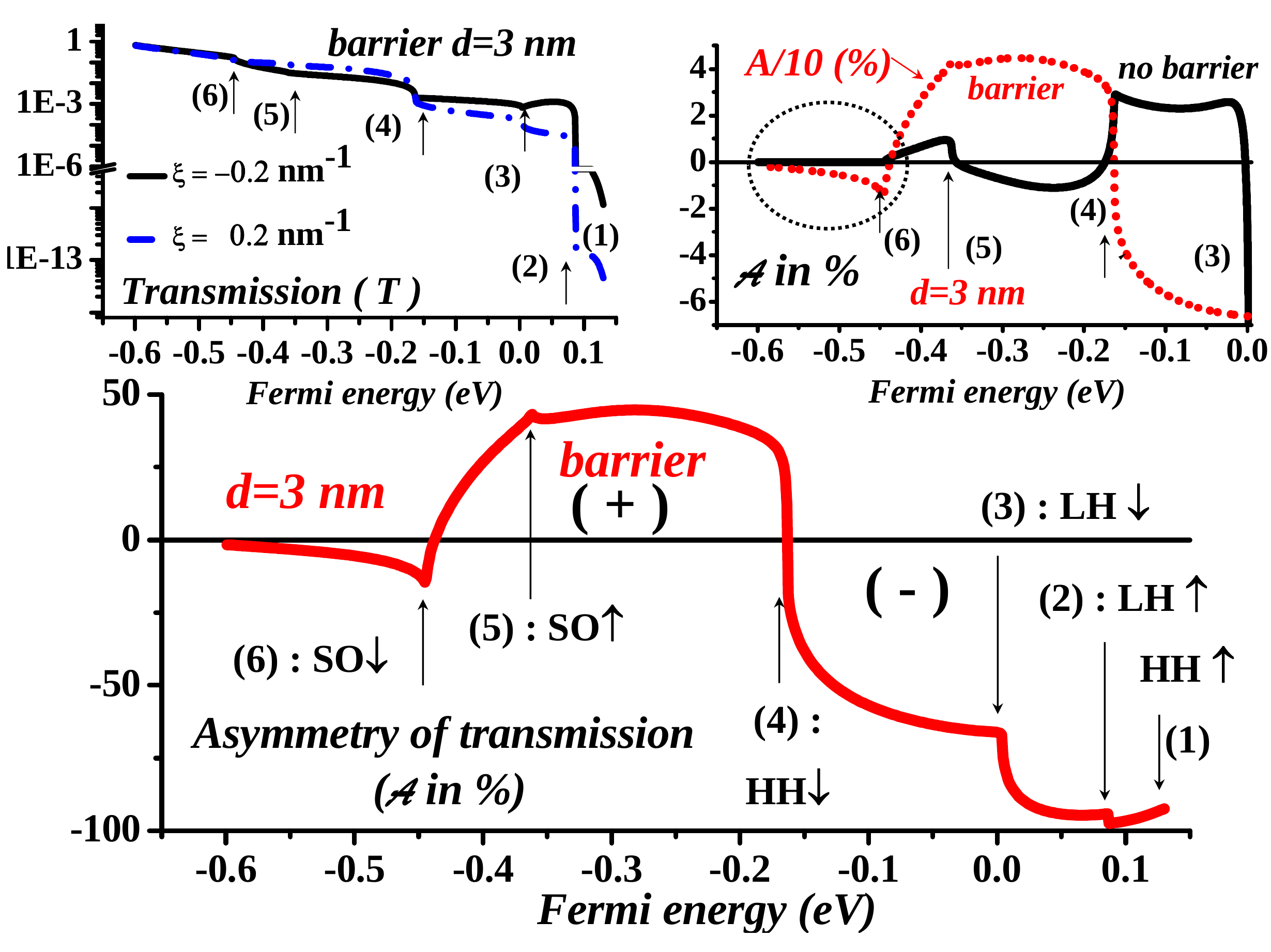}
\caption{(\textit{Bottom}): Transmission asymmetry $\mathcal{A}$ \textit{vs.}
\textit{total} energy $\mathcal{E}$ for a magnetic tunnel junction in the $AP $ state. The parameters are: $2w=0.3$ eV, parallel wavevector $\protect\xi =0.2$~nm$^{-1}$, barrier thickness $d=3$ nm, and barrier height $0.6$ eV. The energy zero corresponds to the non-magnetic upper-valence-band maximum. (\textit{Upper left inset}): transmission $T$ calculated in the $AP$ state; (\textit{Upper right inset}): Asymmetry $\mathcal{A}$ without tunnel barrier ($d=0$; black) compared to the case of the tunnel junction (dotted red).}
\label{fig3}
\end{figure}

To gain physical insight in the case of the valence bands, let us start from a semiconductor toy model with conduction and valence states described by Kane's $S$, $X$, $Y$, and $Z$ real cubic harmonics (no $SOI$)~\cite{kane1957}. Assume that the exchange-split semiconductors are separated by a barrier involving $SOI$, hereafter introduced as a perturbation. Then, the relevant wavefunctions in the barrier are respectively associated to the wavevectors $\xi\widehat{y}\pm i\lambda K\widehat{z}=K\left(t\widehat{y}\pm i\lambda\widehat{z}\right)$ to the left ($+$) and right ($-$). To second order, $\mathbf{k}\cdot\mathbf{p}$ perturbation shows that only the $Y$ and $Z$ valence wavefunctions become coupled through the conduction state, leading to two energy subsets with eigenstates $\mp i\lambda Y+tZ$ (no energy change) and $tY\pm i\lambda Z$ with an energy shift in the gap equal to $\left(\lambda^{2}-t^{2}\right) \left(KP\right)^{2}/E_{G}$, with $P^{2}=|2\gamma_{c}/\hbar \left<S|\hat{p}_{z}|Z\right>|^{2}$. In the $AP$ state, $\left(-i\lambda Y+tZ\right) \uparrow$ to the left [resp. $\left(  tY+i\lambda Z\right) \uparrow$] and $\left(i\lambda Y+tZ\right)  \downarrow$ to the right [resp. $\left(  tY-i\lambda Z\right)  \downarrow$] belong to the same energy shell, resulting in efficient matching at particular incidence $t$. These eigenstates are carrying orbital momentum along $x$, namely $\left\langle L_{x}\right\rangle\ $ proportional to $-\lambda t$, whereas $\left\langle L_{y}\right\rangle =\left\langle L_{z}\right\rangle =0$. Branching $\mathbf{L}\cdot\mathbf{S}$ $SOI$ into the barrier lifts the energy degeneracy between states of same spin associated to $\pm t$ and affects the matching conditions so that asymmetry arises. These effects can be seen as \textit{chirality} phenomena for scattering, analog to magnetic circular dichroism for the optical absorption in ferromagnets. It makes understandable why the association of propagating and evanescent wavevector components may exhalt spinorbitronic effects.

We have presented theoretical evidence for large interfacial scattering asymmetry of carriers \textit{vs.} incidence in semiconducting exchange steps and tunnel barriers. The effect appears to be robust. Direct experimental investigations can probably be performed through angle-resolved photoemission spectroscopy. After averaging over incoming states, a current perpendicular to the barrier is significantly deflected upon tunneling resulting in Tunnel Hall effects and paving the way to original functionalities. Preliminary results indicate that the tunneling asymmetry in the valence band may play an important role in the analysis of tunneling anisotropy magnetoresistance data~\cite{noteSM}.

\vspace{0.1in}

\begin{acknowledgments}
The authors express their gratitude to G. Fishman, J.-M. Jancu, and T. Wade, for  valuable advice. HJ thanks S. Bl\"{u}gel for stimulating discussions. THD acknowledges Idex Paris-Saclay  and Triangle de la Physique for funding. TLHN thanks Nafosted (Grant No. 103.01-2013.25) for support.
\end{acknowledgments}


\begin{thebibliography}{18}
\bibitem{note4} This fact immediatly proves that the effect we describe does not exist in the spin-filter asymmetry considered in Refs.~11 and 12, where the $DP$ spin-orbit Hamiltonian reduces to $\gamma \left( \widehat{\sigma }_{x}k_{x}-\widehat{\sigma}_{y}k_{y}\right) \partial ^{2}/\partial z^{2}$, the term proportional to $\widehat{\sigma}_{z}$ being neglected.
\bibitem{note2} Transport now involves the $D\left( \xi ,k_{1},k_{2}\right) $ coefficient which can be deduced from $C\left( \xi ,k_{1},k_{2}\right) $ by interchanging $k_{1}$ and $k_{2}$ as well as $\xi$ to $-\xi $, resulting in the change of $\mathcal{A}\left( t,\eta \right)$ into $\mathcal{A}\left( -t,\eta \right)=-\mathcal{A}\left( t,\eta \right) $ (this can also be simply checked by symmetry considerations).
\bibitem{note5} It can be checked that the $THE$ voltage is of the order of $(\rho/\delta)L \ell_{mfp} t \mathcal{A} \left(t,\eta \right) J_{z}$, where$L$
is the lateral size of the junction and $\delta$ the thickness of the collecting channel.
\bibitem{note6} The relationship $\eta =\left(-t^{2}+\mathcal{E}/w\right)/\left(1+t^{2}\right)$  relates $\eta$ to the reduced total energy $\mathcal{E}/w$
and the angular averaging is performed with $d\theta=dt/\left(1+t^{2}\right)$.
\end{thebibliography}
\end {document}